\documentclass[runningheads]{llncs}

\usepackage[utf8]{inputenc}
\usepackage{helvet}         
\usepackage{courier}        
\usepackage{type1cm}        
\usepackage{makeidx}         
\usepackage{graphicx}        
\usepackage{multicol}        
\usepackage[bottom]{footmisc}
\usepackage{subcaption}
\usepackage{fancyhdr}
\usepackage{hyperref}
\usepackage[linesnumbered,algoruled,boxed,lined]{algorithm2e}
\usepackage{url}

\begin{document}
\title{Detecting Stable Communities in Link Streams at Multiple Temporal Scales\thanks{This work was supported by the ACADEMICS grant of the IDEXLYON, project of the Universite de Lyon, PIA operated by \textbf{ANR-16-IDEX-0005}, and of the project \textbf{ANR-18-CE23-0004} of the French National Research Agency (ANR).} }
%
%
\author{Souâad Boudebza\inst{1}\and
Rémy Cazabet\inst{2} \and
Omar Nouali\inst{3} \and
Faiçal Azouaou\inst{1}
}
\authorrunning{S. Boudebza et al.}
%

\institute{Ecole nationale Supérieure d'Informatique, BP 68M, 16309, Oued-Smar, Alger, Algeria\\ 
\and
Univ Lyon, Universite Lyon 1, CNRS, LIRIS UMR5205, F-69622 France\\
\and
Division de Recherche en Théorie et Ingénierie des Systèmes Informatiques,CERIST, Rue des Frères Aissiou, Ben Aknoun, Alger, Algeria\\
}
\maketitle              
\begin{abstract}
Link streams model interactions over time in a wide range of fields. Under this model, the challenge is to mine efficiently both temporal and topological structures. Community detection and change point detection are one of the most powerful tools to analyze such evolving interactions. In this paper, we build on both to detect stable community structures by identifying change points within  meaningful communities. Unlike existing dynamic community detection algorithms, the proposed method is able to discover stable communities efficiently at multiple temporal scales. We test the effectiveness of our method on synthetic networks, and on high-resolution time-varying networks of contacts drawn from real social networks.

\end{abstract}
\section{Introduction}
In recent years, studying interactions over time has witnessed a growing interest in a wide range of fields, such as sociology, biology, physics, etc. Such dynamic interactions are often represented using the snapshot model: the network is divided into a sequence of static networks, i.e., snapshots, aggregating all contacts occurring in a given time window. 
The main drawback of this model is that it often requires to choose arbitrarily a temporal scale of analysis.
The link stream model \cite{Latapy2018} is a more effective way for representing interactions over time, that can fully capture the underling temporal information. 

Real world networks evolve frequently at many different time scales. Fluctuations in such networks can be observed at yearly, monthly, daily, hourly, or even smaller scales.  For instance, if one were to look at interactions among workers in a company or laboratory, one could expect to discover clusters of people corresponding to \textit{meetings} and/or \textit{coffee breaks}, interacting at \textbf{high frequency} (e.g., every few seconds) for \textbf{short periods} (e.g., few minutes), \textit{project members} interacting at medium frequency (e.g., once a day) for medium periods (e.g., a few months), \textit{coordination groups} interacting at low frequency (e.g., once a month) for longer periods (e.g., a few years), etc.

An analysis of communities found at an arbitrary chosen scale would necessarily miss some of these communities: low latency ones are invisible using short aggregation windows, while high frequency ones are lost in the noise for long aggregation windows. A multiple temporal scale analysis of communities seems therefore the right solution to study networks of interactions represented as link streams.

To the best of our knowledge, no such method exists in the literature. In this article, we propose a method having roots both in the literature on change point detection and in dynamic community detection. It detects what we call \textbf{stable communities}, i.e., \textit{groups of nodes} forming a \textit{coherent community} throughout a \textit{period of time}, at a given \textit{temporal scale}. 


The remainder of this paper is organized as follows. In Section \ref{related_work}, we present a brief review of related works. Then, we describe the proposed framework in detail in section \ref{framework}. We experimentally evaluate the proposed method on both synthetic and real-world networks in section \ref{experiments}.

\section{Related Work}
\label{related_work}
Our contribution relates to two active body of research: i) dynamic community detection and ii) change point detection. The aim of the former is to discover groups of nodes that persist over time, while the objective of the latter is to detect changes in the overall structure of a dynamic network. In this section, we present existing work in both categories, and how our proposed method relates to them.

\subsection{Dynamic Community Detection}

The problem of detecting communities in dynamic networks has 
attracted a lot of attention in recent years, with various approaches tackling different aspects of the problem, see \cite{rossetti2018community} for a recent survey. Most of these methods consider that the studied dynamic networks are represented as sequences of snapshots, with each snapshot being a well formed graph with meaningful community structure, see for instance \cite{mucha2010community,greene2010tracking}. Some other methods work with interval graphs, and update the community structure at each network change, e.g., \cite{rossetti2017tiles,boudebza2018}.
However, all those methods are not adapted to deal with link streams, for which the network is usually not well formed at any given time. Using them on such a network would require to first aggregate the links of the stream by choosing an arbitrarily temporal scale (aggregation window).


\subsection{Change Point Detection}
Our work is also related to research conducted on change point detection considering community structures. 
In these approaches, given a sequence of snapshots, 
one wants to detect the periods during which the network organization and/or the community structure remains stable. In \cite{Peel204}, the authors proposed the first change-point detection method for evolving networks that uses generative network models and statistical hypothesis testing. 
Wang et al. \cite{Wang2017} proposed a hierarchical change point detection method to detect both inter-community(local change) and intra-community(global change) evolution. 
A recent work 
by Masuda et al. \cite{Masuda2019} used graph distance measures and hierarchical clustering to 
identify sequences of system state dynamics. 

From those methods, our proposal keeps the principle of stable periods delimited by change points, and the idea of detecting changes at local and global scales. But our method differs in two directions: $i)$ we are searching for stable individual communities instead of stable graph periods, and $ii)$ we search for stable structures at multiple levels of temporal granularity.

\section{Method}
\label{framework}
The goal of our proposed method is $i)$ to detect stable communities $ii)$ at multiple scales without redundancy and $iii)$ to do so efficiently. We adopt an iterative approach, searching communities from the coarser to the more detailed temporal scales. At each temporal scale, we use a three step process:
\begin{enumerate}
    \item \textbf{Seed Discovery}, to find relevant community seeds at this temporal scale.
    \item \textbf{Seed Pruning}, to remove seeds which are redundant with communities found at higher scales.
    \item \textbf{Seed Expansion}, expanding seeds in time to discover stable communities.
\end{enumerate}

We start by presenting each of these three steps, and then we describe the method used to iterate through the different scales in section \ref{iterative}.

Our work aims to provide a general framework that could serve as baseline for further work in this field. We define three generic functions that can be set according to the user needs:

\begin{itemize}
    \item \textbf{CD($g$)}, a static community detection algorithm on a graph $g$.
    \item \textbf{QC($N,g$)}, a function to assess the quality of a community defined by the set of nodes $N$ on a graph $g$.
    \item \textbf{CSS($N_1$,$N_2$)}, a function to assess the similarity of two sets of nodes $N_1$ and $N_2$.
\end{itemize}

See section \ref{parameters} on how to choose proper functions for those tasks.

We define a stable dynamic community $c$ as a triplet $c=(N,p,\gamma)$, with $c.N$ the list of nodes in the community, $c.p$ its period of existence defined as an interval, e.g., $c.p=[t_1,t_2[$\footnote{We use right open intervals such as a community starting at $t_x$ and another one ending at the same $t_x$ have an empty intersection, which is necessary to have coherent results when handling discrete time steps.} means that the community $c$ exists from $t_1$ to $t_2$, and $c.\gamma$ the temporal granularity at which $c$ has been discovered. 

We denote the set of all stable dynamic communities $\mathcal{D}$.

\subsection{Seed Discovery}
For each temporal scale, we first search for interesting seeds. A temporal scale is defined by a granularity $\gamma$,  expressed as a period of time (e.g.; 20 minutes, 1 hour, 2 weeks, etc).We use this granularity as a window size, and, starting from a time $t_0$ --by default, the date of the first observed interaction-- we create a cumulative graph (snapshot) for every period $[t_0,t_0+\gamma[,[t_0+\gamma,t_0+2\gamma[,[t_0+2\gamma,t_0+3\gamma[,etc.$, until all interactions belong to a cumulative graph. This process yields a sequence of static graphs, such as $G_{t_0,\gamma}$ is a cumulated snapshot of link stream $G$ for the period starting at $t_0$ and of duration $\gamma$. $G_{\gamma}$ is the list of all such graphs.

Given a static community detection algorithm $CD$ yielding a set of communities, and a function to assess the quality of communities $QC$, we 
apply $CD$ on each snapshot and filter promising seeds, i.e., high quality communities,
using $QC$. The set of valid seeds $\mathcal{S}$ is therefore defined as: 
 
\begin{equation} 
\mathcal{S} = \{\forall g \in G_{\gamma}, \forall s \in CD(g), QC(s,g)>\theta_q\}
\end{equation}
With $\theta_q$ a threshold of community quality.  

Since community detection at each step is independent, we can run it in parallel on all steps, this is an important aspect for scalability. 

\subsection{Seed Pruning}
The seed pruning step has a twofold objective: $i)$ reducing redundancy and $ii)$ speed up the multi-scale community detection process. Given a measure of structural similarity $CSS$, we prune the less interesting seeds, such as the set of filtered seeds $\mathcal{FS}$ is defined as:

\begin{equation} 
\mathcal{FS} = \{\forall s \in \mathcal{S}, \forall c \in \mathcal{D}, (CSS(s.N, c.N)>\theta_s) \vee (s.p \cap c.p = \{\emptyset\})
\end{equation}

Where $\mathcal{D}$ is the set of stable communities discovered at coarser (or similar, see next section) scales, $s.p$ is the interval corresponding to the snapshot at which this seed has been discovered, and $\theta_s$ is a threshold of similarity.

Said otherwise, we keep as interesting seeds those that are not redundant topologically (in term of nodes/edges), OR not redundant temporally. A seed is kept if it corresponds to a situation never seen before.

\subsection{Seed Expansion}
The aim of this step is to assess whether a seed corresponds to a \textit{stable} dynamic community.
The \textit{instability} problem has been identified since the early stages of the dynamic community detection field \cite{aynaud2010static}. It means that the same algorithm ran twice on the same network after introducing minor random modifications might yield very different results. As a consequence, one cannot know if the differences observed between the community structure found at $t$ and at $t+1$ are due to structural changes or to the instability of the algorithm. This problem is usually solved by introducing smoothing techniques \cite{rossetti2018community}. Our method use a similar approach, but 
instead of comparing communities found at step $t$ and $t-1$, we check whether a community found at $t$ is still relevant in previous and following steps, recursively. 



More formally, for each seed $s \in FS$ found on the graph $G_{t,\gamma}$, we iteratively expand the duration of the seed $s.d=[t,t+\gamma[$ (where $t$ is the time start of this duration) at each step $t_i$ in both temporal directions ($t_i \in (...[t-2\gamma,t-\gamma[,[t-\gamma,t]; [t+\gamma,t+2\gamma[,[t+2\gamma,t+3\gamma]...))$ as long as the quality $QC(s.N,G_{t_i,\gamma})$ of the community defined by the nodes $s.N$ on the graph at $G_{t_i,\gamma}$ is good enough.
Here, we use the same similarity threshold $\theta_s$ as in the seed pruning step.
If the final duration period $|s.p|$ of the expanded seed is higher than a duration $\theta_p \gamma$, with $\theta_p$ a threshold of stability, the expanded seed is added to the list of stable communities, otherwise, it is discarded. This step is formalized in Algorithm \ref{alg:extend}.

\begin{algorithm}[H]
\SetAlgoLined
\KwIn{$s, \gamma,\theta_p,\theta_s 
$}
 $t \gets t^{start} | s.p = [t^{start},t^{end}[$ \;
 $g \gets G_{t,\gamma}$\;
 $p \gets [t,t+\gamma[$\;
 \While{$QC(s.N,g)>\theta_s$}{
  $s.p \gets s.p \cup p$\;
  $t \gets t+\gamma $\;
  $p \gets [t,t+\gamma[$\;
  $g \gets G_{t,\gamma}$\;
  }
  \If{$|s.p| \geq \theta_p \gamma$}{
    $\mathcal{D} \gets  \mathcal{D} \cup \{s$\}\;
   }

 \caption{\textbf{Forward seed expansion}. 
 Forward temporal expansion of a seed $s$ found at time $t$ of granularity  $\gamma$. The reciprocal algorithm is used for \textit{backward} expansion: $t+1$ becomes $t-1$.}
 \label{alg:extend}
\end{algorithm}

In order to select the most relevant stable communities, 
we consider seeds in descending order of their $QC$ score, i.e., the seeds of higher quality scores are considered first.
Due to the pruning strategy, a community of lowest quality might be pruned by a community of highest quality at the same granularity $\gamma$.

\subsection{Multi-scale Iterative Process}
\label{iterative}
Until then, we have seen how communities are found for a particular time scale. In order to detect communities at multiple scales, we first define the ordered list of studied scales $\Gamma$. The largest scale is defined as $\gamma^{max}=|G.d|/\theta_p$, with $|G.d|$ the total duration of the dynamic graph. Since we need to observe at least $\theta_p$ successive steps to consider the community stable, $\gamma^{max}$ is the largest scale at which communities can be found.

We then define $\Gamma$ as the ordered list: 

\begin{equation}
    \Gamma=[\gamma^{max}, \gamma^{max}/2^1, \gamma^{max}/2^2, \gamma^{max}/2^3,..., \gamma^{max}/2^k ] 
\end{equation}
With $k$ such as $\gamma^{max}/2^k>\theta_{\gamma}>= \gamma^{max}/2^{k+1} $, $\theta_{\gamma}$ being a parameter corresponding to the finest temporal granularity to evaluate, which is necessarily data-dependant (if time is represented as a continuous property, this value can be fixed at least at the sampling rate of data collection).

This exponential reduction in the studied scale guarantees a limited number of scales to study.

The process to find seeds and extend them into communities is then summarized in Algorithm \ref{alg:iterative}.

\begin{algorithm}[H]
\SetAlgoLined
\KwIn{$G,\theta_q,\theta_s,\theta_p,\theta_{\gamma}$}
$\mathcal{D} \gets \{\emptyset\}$\;
$\Gamma \gets $studied\_scales($G,\theta_{\gamma})$ \;
 \For{$\gamma \in \Gamma$}{
    $\mathcal{S} \gets $ Seed\_Discovery($\gamma
    ,CD,QC,\theta_q$)\;
    $\mathcal{FS} \gets $Seed\_Pruning($\mathcal{S},CSS,\theta_s$)\;
    \For{$s \in \mathcal{FS}$}{
        Seed\_Expansion($s,\gamma,\theta_p,\theta_s$)\;
     }
  }

 \caption{\textbf{Multi-temporal-scale stable communities finding}. Summary of the proposed method. See corresponding sections for the details of each step. $G$ is the link streams to analyze, $\theta_q,\theta_s,\theta_p,\theta_{\gamma}$ are threshold parameters.
}
 \label{alg:iterative}
\end{algorithm}

\subsection{Choosing Functions and Parameters}
\label{parameters}
The proposed method is a general framework that can be implemented using different functions for $CD, QC$ and $CSS$. 
This section provides explicit guidance for selecting each function, and introduces the choices we make for the experimental section. 

\subsubsection{Community Detection - CD}
Any algorithm for community detection could be used, including overlapping methods, since each community is considered as an independant seed. Following literature consensus, we use the Louvain method \cite{blondel2008fast}, which yields non-overlapping communities using a greedy modularity-maximization method. The louvain method performs well on static networks, it is in particular among the fastest and most efficient methods.
Note that it would be meaningful to adopt an algorithm yielding communities of good quality according to the chosen $QC$, which is not the case in our experiments, as we wanted to use the most standard algorithms and quality functions in order to show the genericity of our approach.

\subsubsection{Quality of Communities - QC}
The $QC$ quality function must express the quality of a set of nodes w.r.t a given network, unlike functions such as the modularity, which express the quality of a whole partition w.r.t a given network. Many such functions exist, like \textit{Link Density} or \textit{Scaled Density}\cite{Labatut2017}, but the most studied one is probably the \textit{Conductance} \cite{leskovec2009community}. Conductance 
is defined as the ratio of i)the number of edges between nodes inside the community and nodes outside the community, and ii)the sum of degrees of nodes inside the community (or outside, if this value is larger). More formally, the conductance $\phi$ of a community $C$ is :
\[\phi(C)=\frac{\sum_{i\in C, j \notin C} A_{i,j}}{Min (A(C), A(\bar{C}))}\]
Where $A$ is the adjacency matrix of the network, \(A(C)=\sum_{i \in C}\sum_{j \in V}A_{i,j}\) and $\bar{C}$ is the complement of $C$. Its value ranges from 0 (Best, all edges starting from nodes of the community are internal) to 1 (Worst, no edges between this community and the rest of the network). 
Since our generic framework expects good communities to have $QC$ scores higher than the threshold $\theta_q$, we adopt the definition $QC$=1-conductance. 

\subsubsection{Community Seed Similarity - CSS}
This function takes as input two sets of nodes, and returns their similarity. Such a function is often used in dynamic community detection to assess the similarity between communities found in different time steps. Following \cite{greene2010tracking}, we choose as a reference function the Jaccard Index.
Given two sets A and B, it is defined as: $J(A,B) = \frac{|A \cap B|}{ |A \cup B|}$

\subsection{Parameters}
The algorithm has four parameters, $\theta_{\gamma},\theta_q,\theta_s,\theta_p$, defining different thresholds. We explicit them and provide the values used in the experiments.  
\begin{enumerate}
    \item $\theta_{\gamma}$ is data-dependant. It corresponds to the smallest temporal scale that will be studied, and should be set at least at the collection rate. 
    For synthetic networks, it is set at 1 (the smallest temporal unit needed to generate a new stream), while, for SocioPatterns dataset, it is set to 20 secondes(the minimum length of time required to capture a contact). 
    \item $\theta_q$ determines the minimal quality a seed must have to be preserved and expanded. The higher this value, the more strict we are on the quality of communities. We set $\theta_q=0.7$ in all experiments. It is dependent on the choice of the $QC$ function.
    \item $\theta_s$ determines the threshold above which two communities are considered redundant. The higher this value, the more communities will be obtained. We set $\theta_s=0.3$ in all experiments. It is dependent on the choice of the $CSS$ function.
    \item $\theta_p$ is the minimum number of consecutive periods a seed must be 
    expanded in order to be considered as stable community. We set $\theta_s=3$ in all experiments. The value should not be lower in order to avoid spurious detections due to pure chance. Higher values could be used to limit the number of results.
\end{enumerate}

\section{Experiments and Results}
\begin{figure}[!h]
\centering
\begin{subfigure}{0.8\textwidth}
    \includegraphics[width=\textwidth]{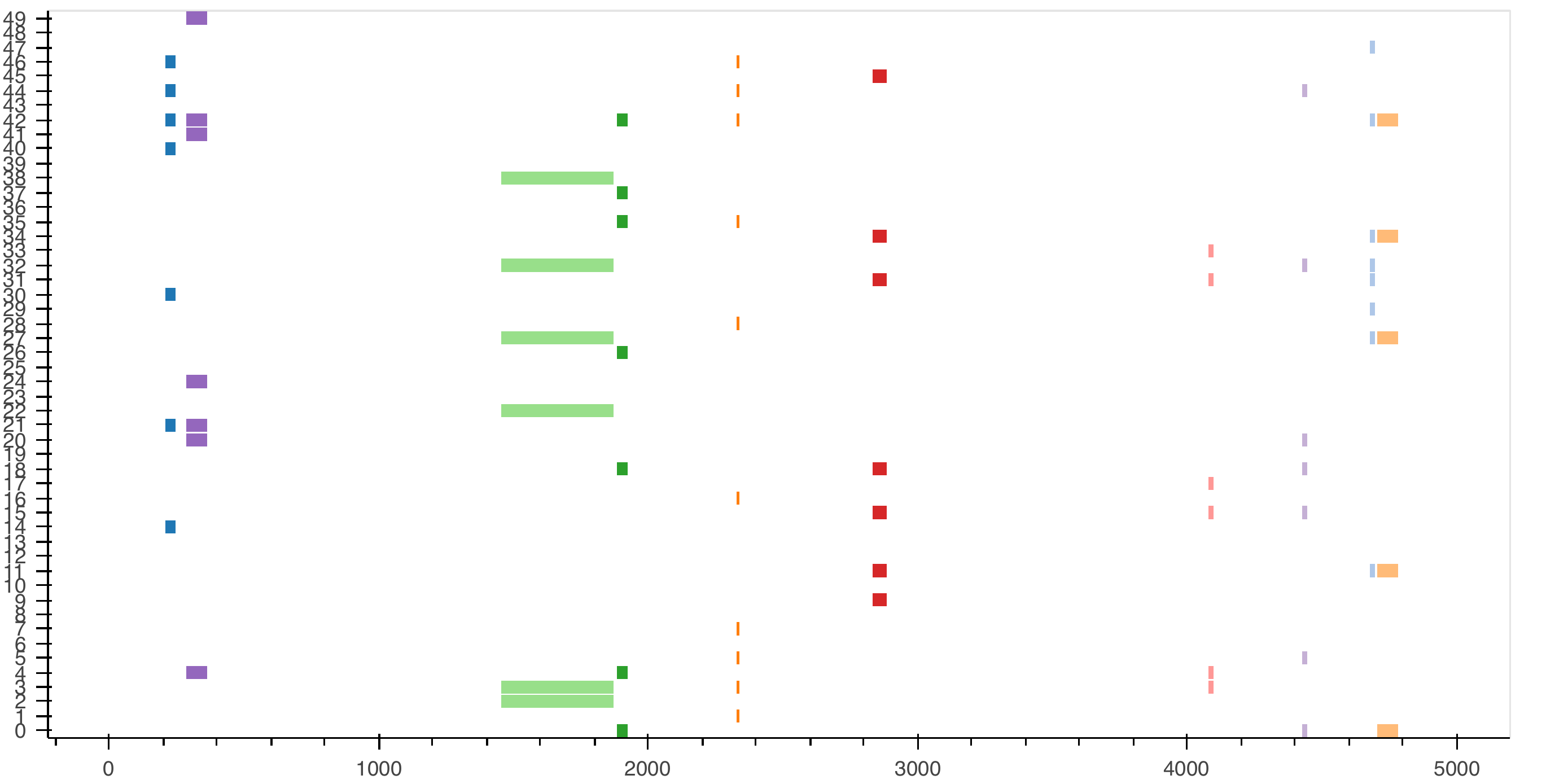}
    \caption{Stable communities produced by the generator.}
\end{subfigure}

\begin{subfigure}{0.8\textwidth}
    \includegraphics[width=\textwidth]{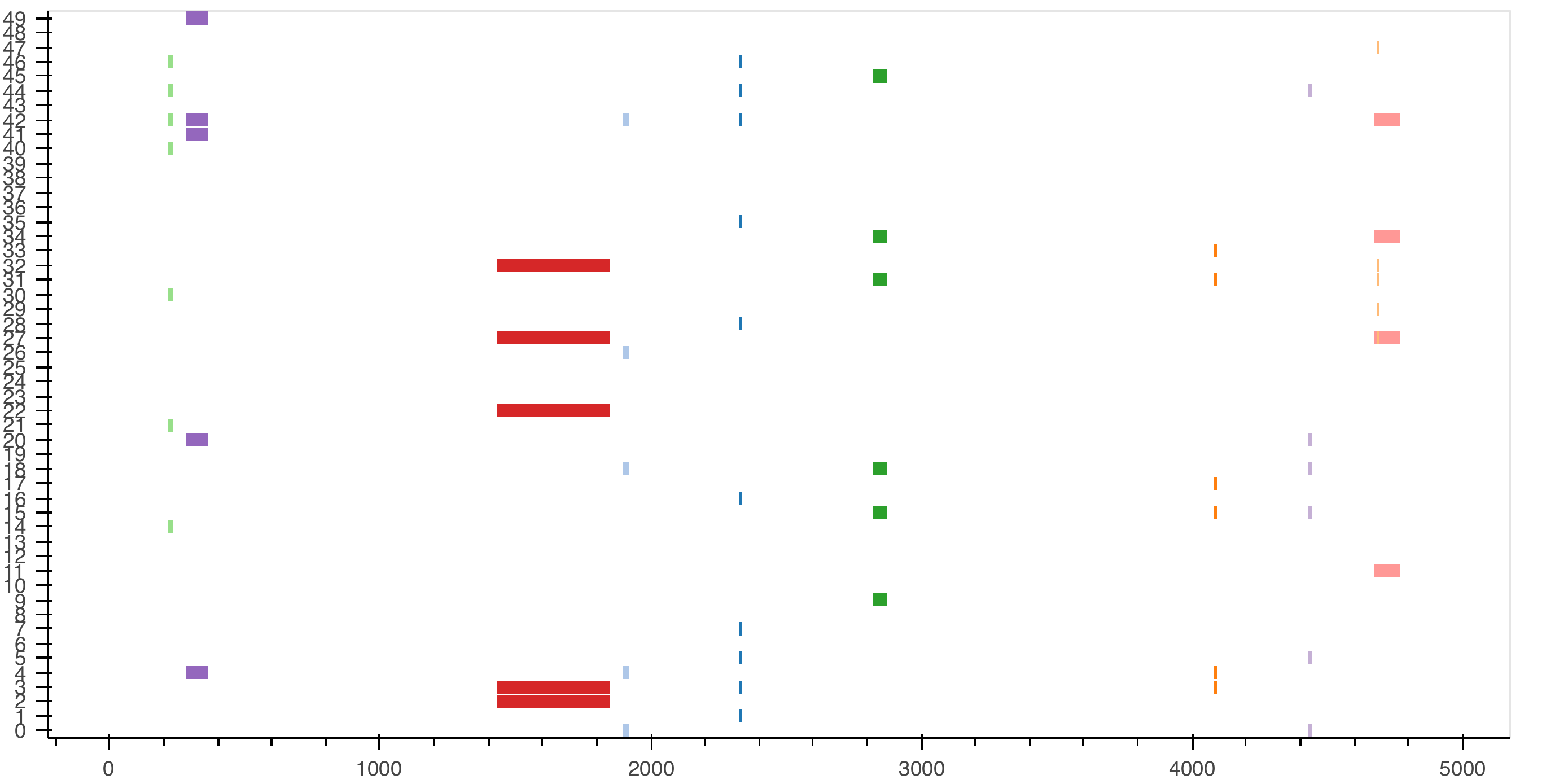}
        \caption{Stable communities discovered by the proposed method.}
\end{subfigure}

	\caption{Visual comparison between planted and discovered communities. Time steps on the horizontal axis, nodes on the vertical axis.  Colors correspond to communities and are randomly assigned. We can observe that most communities are correctly discovered, both in terms of nodes and of duration.}
	\label{fig:synthetic}	
\end{figure}

\label{experiments}


The validation of our method encompasses 
three main aspects: $i)$ the validity of communities found, and $ii)$ the multi-scale aspect of our method, $iii)$ its scalability. We conduct two kinds of experiments: on synthetic data, on which we use planted \textit{ground-truth} to quantitatively compare our results, and on real networks, on which we use both qualitative and quantitative evaluation to validate our method.  

\subsection{Validation on Synthetic Data}
To the best of our knowledge, no existing network generator allows to generate dynamic communities at multiple temporal scale. We therefore introduce a simple solution to do so.
Let us consider a dynamic network composed of $T$ steps and $N$ different nodes. We start by adding some random noise: at each step, an Erdos-Renyi random graph\cite{erdos1959} is generated, with a probability of edge presence equal to $p$. We then add a number $SC$ of random stable communities. For each community, we attribute randomly a set of $n\in [4,N/4]$ nodes, a duration $d \in [10,T/4]$ and a starting date $s\in [0,T-d]$. $n$ and $d$ are chosen using a logarithmic probability, in order to increase variability. The temporal scale of the community is determined by the probability of observing an edge between any two of its nodes during the period of its existence, set as $10/d$. As a consequence, a community of duration 10 will have edges between all of its nodes at every step of its existence, while a community of length 100 will have an edge between any two of its nodes only every 10 steps in average.

Since no algorithm exists to detect communities at multiple temporal scales, we compare our solution to a baseline: communities found by a static algorithm on each window, for different window sizes. It corresponds to \textit{detect \& match}
methods for dynamic community detection such as \cite{greene2010tracking}. We then compare the results by computing the overlapping NMI as defined in \cite{lancichinetti2009detecting}, at each step. For those experiments, we set $T=5000,N=100,p=10/N$. We vary the number of communities $SC$.

\begin{table}[]
\centering
\begin{tabular*}{\textwidth}{c @{\extracolsep{\fill}}clcccccc}
 t\_scale ($\gamma$) &    5 &   10 &   20 &   30 &   40 &   50 \\ \hline
     \textit{Proposed} & \textbf{0.91} & \textbf{0.78} & \textbf{0.69} & \textbf{0.69} & \textbf{0.62} & \textbf{0.54} \\
    1666 & 0.41 & 0.32 & 0.24 & 0.23 & 0.15 & 0.19 \\
     833 & 0.36 & 0.30 & 0.29 & 0.27 & 0.23 & 0.25 \\
    416 & 0.39 & 0.40 & 0.36 & 0.34 & 0.32 & 0.33 \\
    208 & 0.46 & 0.45 & 0.40 & 0.42 & 0.41 & 0.37 \\
     104 & 0.47 & 0.48 & 0.44 & 0.46 & 0.45 & 0.42 \\
   52 & 0.45 & 0.47 & 0.45 & 0.47 & 0.47 & 0.45 \\
    26 & 0.35 & 0.35 & 0.38 & 0.42 & 0.42 & 0.41 \\
     13 & 0.28 & 0.26 & 0.30 & 0.31 & 0.32 & 0.31 \\
     6 & 0.17 & 0.16 & 0.19 & 0.19 & 0.20 & 0.19 \\
      3 & 0.12 & 0.09 & 0.11 & 0.10 & 0.12 & 0.11 \\
      1 & 0.05 & 0.03 & 0.04 & 0.03 & 0.05 & 0.04 \\
\end{tabular*}
\caption{Comparison of the Average NMI scores(over 10 runs) obtained for the proposed method (\textit{Proposed}) and for each of the temporal scales ($\gamma \in \Gamma$) used by the proposed method, taken independently.}
\label{tab:averagenmi}
\end{table}

Figure \ref{fig:synthetic} represents the synthetic communities to find for $SC=10$, and the communities discovered by the proposed method. We can observe a good match, with communities discovered throughout multiple scales (short-lasting and long-lasting ones).
We report the results of the comparison with baselines in table \ref{tab:averagenmi}. We can observe that the proposed method outperforms the baseline at every scale in all cases in term of average NMI.

The important implication is that the problem of dynamic community detection is not only a question of choosing the right scale through a window size, but that if the network contains communities at multiple temporal scale, one needs to use an adapted method to discover them.

\subsection{Validation on Real Datasets}

\begin{figure}[!h]
\centering
    \begin{subfigure}{0.7\textwidth}
        \includegraphics[width=\textwidth]{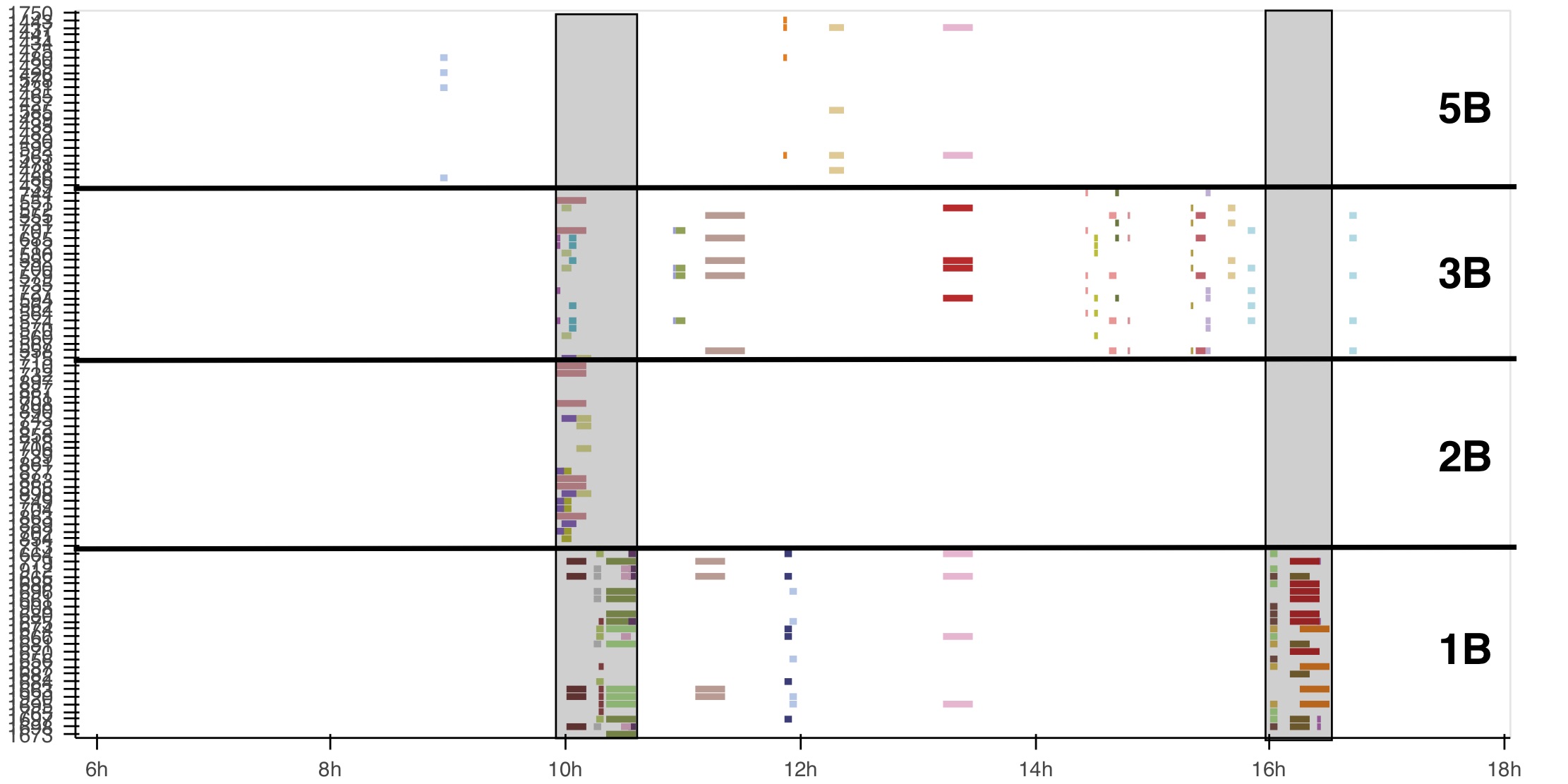}
        \caption{Second day, length$<$30min. Grey vertical areas correspond to most likely break periods.}
        \label{fig:AllComs}
    \end{subfigure}
    
    \begin{subfigure}{0.7\textwidth}
        \includegraphics[width=\textwidth]{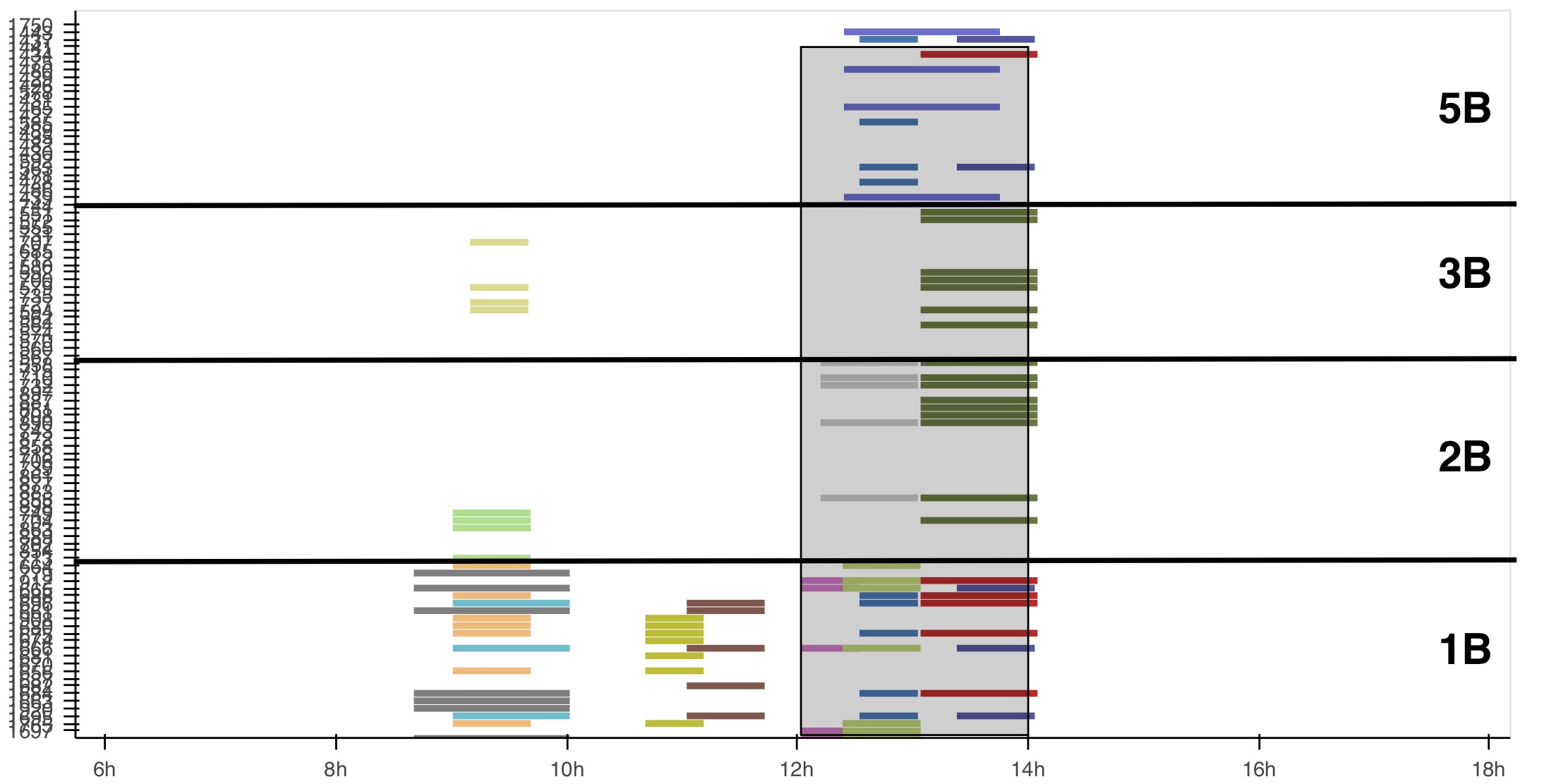}
        \caption{Second day, 
        30min$<$length$<$2hours. Grey vertical area corresponds to the lunch break}
        \label{fig:TwentyComs}
    \end{subfigure}
    
    \begin{subfigure}{0.7\textwidth}
        \includegraphics[width=\textwidth]{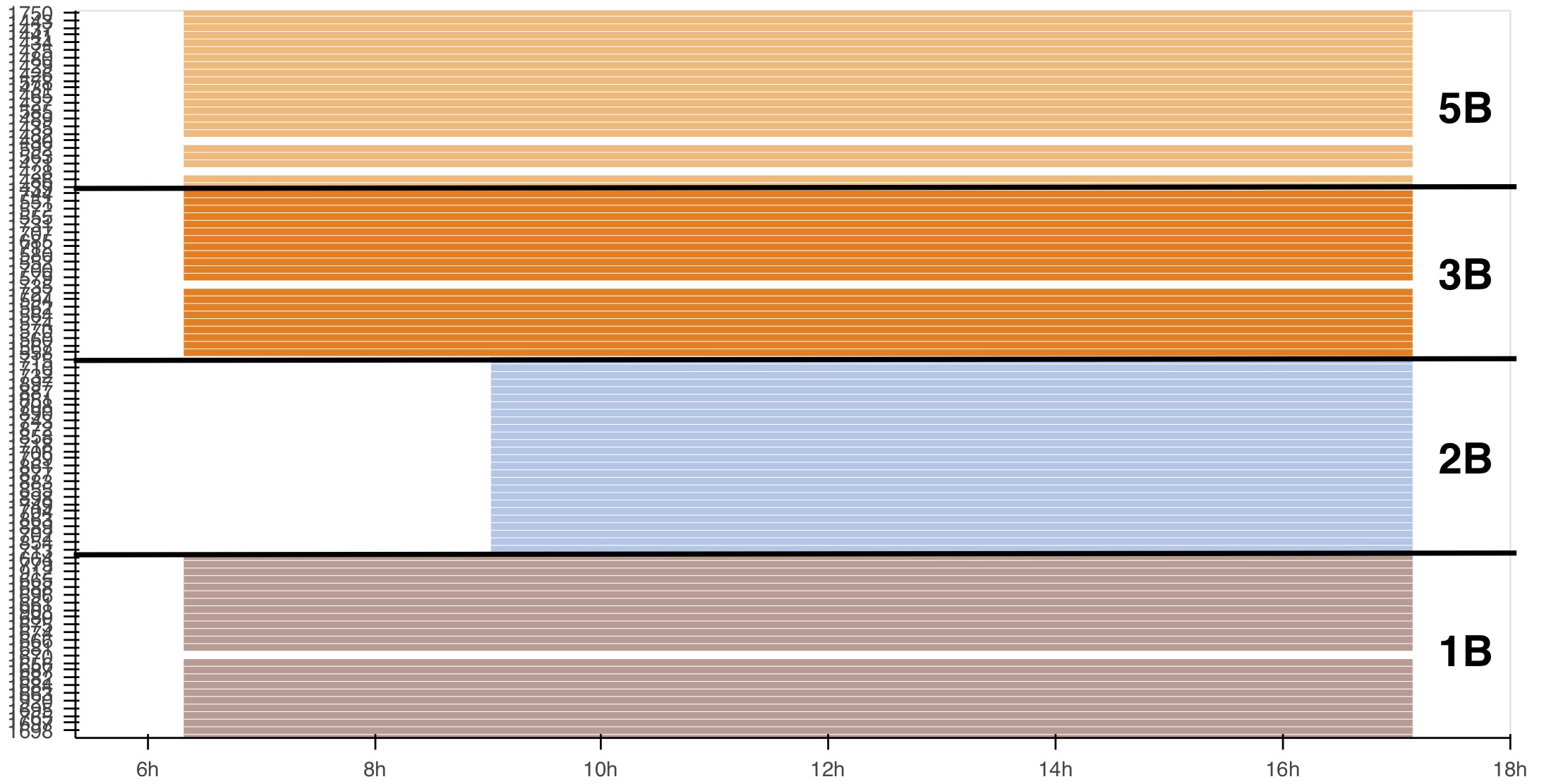}
        \caption{Second day, length$>$2hours}
        \label{fig:SecondDayComs}
    \end{subfigure}    
	\caption{Stable communities of different lengths on the SocioPatterns Primary School Dataset. Time on the horizontal axis, children on the vertical axis. Colors are attributed randomly.}
	\label{fig:PrimarySchool}	
\end{figure} 


We validate our approach by applying it to two 
real datasets. Because no ground truth data exists to compare our results with, we validate our method by using both quantitative and qualitative evaluation. We use the quantitative approach to analyze the scalability of the method and the characteristics of communities discovered compared with other existing algorithms. We use the qualitative approach to show that the communities found are meaningful and could allow an analyst to uncover interesting patterns in a dynamic datasets.

The datasets used are the following:
\begin{itemize}
    \item \textbf{SocioPatterns} primary school data\cite{sociopattrens2011}, face-to-face interactions between children in a school (323 nodes, 125 773 interaction). 
    \item \textbf{Math overflow} stack exchange interaction dataset \cite{paranjape2017motifs}, a larger network to evaluate scalability (24 818 nodes, 506 550 interactions).
\end{itemize}{}

\subsubsection{Qualitative evaluation}
For the qualitative evaluation, we used the primary school data\cite{sociopattrens2011} collected by the SocioPatterns collaboration \footnote{www.sociopatterns.org} using RFID devices. They capture face-to-face proximity of individuals wearing them, at a rate of one capture every 20 seconds. The dataset contains face-to-face interactions between 323 children and 10 teachers collected over two consecutive days in October 2009. 
This school has 5 levels, each level is divided into 2 classes(A and B), for a total of 10 classes.

No community ground truth data exists to validate quantitatively our findings. We therefore focus on the descriptive information highlighted on the SocioPatterns study \cite{sociopattrens2011}, and we show how the results yielded by our method match the course of the day as recorded by the authors in this study.

In order to make an accurate analysis of our results, the visualization have been reduced to one day (the second day), and we limited ourselves to 4 classes (1B, 2B, 3B, 5B)
\footnote{Note that full results can be explored online using the provided notebook (see conclusion section)}. 
120 communities are discovered in total on this dataset. We created three different figures, corresponding to communities of length respectively i)less than half an hour, ii) between half an hour and 2 hours, iii) more than 2 hours. Figure \ref{fig:PrimarySchool} depicts the results. Nodes affiliations are ordered by class, as marked on the right side of the figure. The following observations can be made:
\begin{itemize}
    \item Communities having the longest period of existence clearly correspond to the class structure. Similar communities had been found by the authors of the original study using aggregated networks per day. 
    \item Most communities of the shorter duration are detected during what are probably breaks between classes. In the original study, it had been noted that break periods are marked by the highest interaction rates
    . We know from data description that classes have 20/30 minutes breaks, and that those breaks are not necessarily synchronized between classes. This is compatible with observation, in particular with communities found between 10:00 and 10:30 in the morning, and between 4:00 and 4:30 in the afternoon.
    \item Most communities of medium duration occur during the lunch break. We can also observe that the most communities are separated into two intervals, 12:00-13:00 and 13:00-14:00.  This can be explained by the fact that children have a common canteen, and a shared playground. As the playground and the canteen do not have enough capacity to host all the students at the same time, only two or three classes have breaks at the same time, and lunches are taken in two consecutive turns of one hour. Some children do not belong to any communities during the lunch period, which matches the information that about half of the children come back home for lunch \cite{sociopattrens2011}.
    \item During lunch breaks and class breaks, some communities involve children from different classes, see the community with dark-green colour during lunch time (medium duration figure) or the pink community around 10:00 for short communities, when classes 2B and 3B are probably in break at the same time. This confirms that an analysis at 
    coarser scales only can be misleading, as it leads only to the detection of the stronger class structure, ignoring that communities exist between classes too, during shorter periods.
\end{itemize}

\subsubsection{Quantitative evaluation}
In this section, we compare our proposition with other methods on two aspects: scalability, and aggregated properties of communities found. The methods we compare ourselves to are: 
\begin{itemize}
    \item An Identify and Match framework proposed by Greene et al. \cite{greene2010tracking}. We implement it using the Louvain method for community detection, and the  \textit{Jaccard coefficient} to match communities, with a minimal similarity threshold of 0.7. We used a custom implementation, sharing the community detection phase with our method.
    \item The multislice method introduced by Mucha et al. \cite{mucha2010community}. We used the authors implementation, with interslice coupling $\omega=0.5$.
    \item The dynamic clique percolation method (D-CPM) introduced by Palla et al. \cite{palla2007quantifying}. We used a custom implementation, the detection in each snapshot is done using the implementation in the networkx library \cite{hagberg2008exploring}.
\end{itemize}
For Identify and Match, D-CPM and our approach, the community detection phase is performed in parallel for all snapshots. This is not possible for Mucha et al., since the method is performed on all snapshots simultaneously. On the other hand, D-CPM and Indentify and Match are methods with no dynamic smoothing.

\begin{figure}[!h]
\centering
    \includegraphics[width=0.7\textwidth]{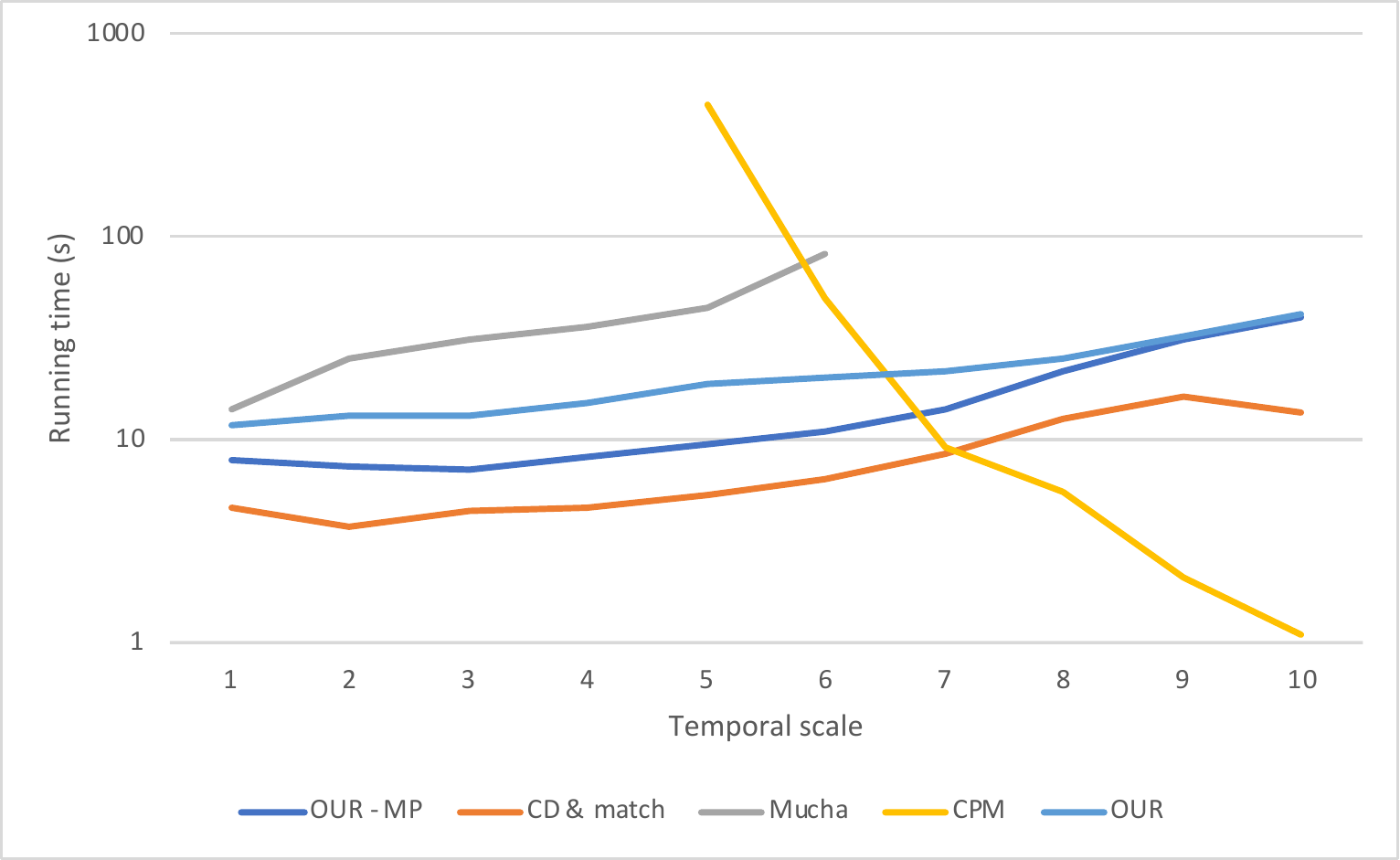}

	\caption{Speed of several dynamic community detection methods for several temporal granularities, on the Math Overflow dataset. Missing points correspond to computation time above 1000s. Temporal scales correspond to window sizes and are divided by 2 at every level, from 1=67 681 200s (about 2 years) to 10=132 189s (about 36h). OUR and OUR-MP corresponds to our method using or not multiprocessing (4 cores)}
	\label{fig:scalability}	
\end{figure}

Figure \ref{fig:scalability} presents the time taken by those methods and our proposition, for each temporal granularity, on the Math Overflow network. The task accomplished by our method is, of course, not comparable, since it must not only discover communities, but also avoid redundancy between communities in different temporal scales, while other methods yield redundant communities in different levels. Nevertheless, we can observe that the method is scalable to networks with tens of thousands of nodes and hundreds of thousands of interactions. It is slower than the Identify and Match(CD\&Match) approach, but does not suffer from the scalability problem as for the two other ones(D-CPM and Mucha et al.,). In particular, the clique percolation method is not scalable to large and dense networks, a known problem due to the exponential growth in the number of cliques to find. For the method by Mucha et al., the scalability issue is due to the memory representation of a single modularity matrix for all snapshots.



\begin{table}[]
\centering
\begin{tabular}{lllllll}
Method   & \#Communities & Persistance & Size  & Stability & Density & Q \\
\hline
OUR      & 179 & 3.44        & 10.89 & 1.00      & 0.50    & 0.91        \\
CD\&MATCH & 29846 & 1.21        & 5.50  & 0.97      & 0.42    & 0.96        \\
CPM     & 3259  & 1.87        & 5.37  & 0.51      & 0.01    & 0.53        \\
MUCHA   & 1097  & 15.48       & 9.72  & 0.62      & 0.38    & 0.85       
\end{tabular}
\caption{Average properties of communities found by each method (independently of their temporal granularity). \#Communities: number of communities found. Persistence: number of consecutive snapshots. Size: number of nodes. Stability: average Jaccard coefficient between nodes of the same community in successive snapshots. Density: average degree/size-1. Q: 1-Conductance (higher is better)}
\label{tab:carac}
\end{table}

In table \ref{tab:carac}, we summarize the number of communities found by each method, their persistence, size, stability, density and conductance. It is not possible to formally rank those methods based on these values only, that correspond to vastly different scenarios. What we can observe is that existing methods yield much more communities than the method we propose, usually at the cost of lower overall quality. 
When digging into the results, it is clear that other methods yield many noisy communities, either found on a single snapshot for methods without smoothing, unstable for the smoothed Mucha method, and often with low density or Q.
\section{Conclusion and future work}
To conclude, this article only scratches the surface of the possibilities of multiple-temporal-scale community detection. We have proposed a first method for the detection of such structures, that we validated on both synthetic and real-world networks, highlighting the interest of such an approach. The method is proposed as a general, extensible framework, and its code is available \footnote{The full code is available at \url{https://github.com/Yquetzal/ECML_PKDD_2019}}\footnote{An online notebook to test the method is available at \url{https://colab.research.google.com/github/Yquetzal/ECML_PKDD_2019/blob/master/simple_demo.ipynb}}as an easy to use library, for replications, applications and extensions.

As an exploratory work, further investigations and improvements are needed. Heuristics or statistical selection procedures could be implemented to reduce the computational complexity. Hierarchical organization of relations --both temporal and structural--between communities could greatly simplify the interpretation of results. 

%
%
%
\bibliographystyle{splncs04}
\bibliography{ref}
\end{document}